\documentclass[twocolumn,showpacs,preprintnumbers,amsmath,amssymb]{revtex4-1}

\usepackage{graphicx}
\usepackage{dcolumn}
\usepackage{bm}
\usepackage{epstopdf}
\usepackage{amssymb}

\usepackage{epsfig}
\usepackage{tikz}        
\usepackage{etoolbox}

\usepackage{lineno}
\usepackage[caption=false]{subfig} 

\begin{document}

\newrobustcmd*{\mydiamond}[1]{\tikz{\filldraw[black,fill=#1] 
(0,0) -- (0.2cm,0.2cm) -- (0.4cm,0) -- (0.2cm,-0.2cm) -- (0,0);}}

\newrobustcmd*{\mytriangleright}[1]{\tikz{\filldraw[black,fill=#1] (0,0.2cm) -- 
(0.3cm,0) -- (0,-0.2cm);}}


\title{Beyond the ``faster is slower'' effect} %

 \author{I.M.~Sticco}
 \affiliation{Departamento de F\'\i sica, Facultad de Ciencias Exactas y 
Naturales, Universidad de Buenos Aires,\\
 Pabell\'on I, Ciudad Universitaria, 1428 Buenos Aires, Argentina.}
 \author{F.E.~Cornes}
 \affiliation{Departamento de F\'\i sica, Facultad de Ciencias Exactas y 
Naturales, Universidad de Buenos Aires,\\
 Pabell\'on I, Ciudad Universitaria, 1428 Buenos Aires, Argentina.}
\author{G.A.~Frank}
 \affiliation{Unidad de Investigaci\'on y Desarrollo de las 
Ingenier\'\i as, Universidad Tecnol\'ogica Nacional, Facultad Regional Buenos 
Aires, Av. Medrano 951, 1179 Buenos Aires, Argentina.}
\author{C.O.~Dorso}%
 \email{codorso@df.uba.ar}
\affiliation{Departamento de F\'\i sica, Facultad de Ciencias Exactas y 
Naturales, Universidad de Buenos Aires,\\
 Pabell\'on I, Ciudad Universitaria, 1428 Buenos Aires, Argentina.}%
\affiliation{Instituto de F\'\i sica de Buenos Aires, Pabell\'on I, 
Ciudad Universitaria, 1428 Buenos Aires, Argentina.}
 
\date{\today}

\begin{abstract}
The ``faster is slower'' effect raises when crowded people push each other to 
escape through an exit during an emergency situation. As individuals push 
harder, a statistical slowing down in the evacuation time can be achieved. The 
slowing down is caused by the presence of small groups of pedestrians (say, a 
small human cluster) that temporary blocks the way out when trying to leave the 
room. The pressure on the pedestrians belonging to this blocking cluster raises 
for increasing anxiety levels and/or larger number of individuals trying to 
leave the room through the same door. Our investigation shows, however, that 
very high pressures alters the dynamics in the blocking cluster, and thus, 
changes the statistics of the time delays along the escaping process. It can be 
acknowledged a reduction in the long lasting delays, while the overall 
evacuation performance improves. We present results on this novel phenomenon 
taking place beyond the ``faster is slower'' regime. 
\end{abstract}


\maketitle

\section{\label{introduction}Introduction}

The ``faster is slower'' (FIS) effect is the major phenomenon taking place when 
pedestrians get involved in a dangerous situation and try to escape through an 
emergency door. It states that the faster they try to reach the exit, the 
slower they move due to clogging near the door. This effect has been 
observed in the context of the ``social force model'' (SFM) \cite{Helbing1}. 
But research on other physical systems, such as grains flowing out a 2D hopper 
or sheep entering a barn, are also known to exhibit a ``faster is slower'' 
behavior \cite{Parisi}. \\

Research on the clogging delays (in the context of the SFM) has shown 
that a small group of pedestrians close to the door is responsible 
for blocking the way to the rest of the crowd. This \textit{blocking 
clusters} appear as an arch-like metastable structure 
around the exit. The tangential friction between pedestrians belonging to this
blocking structure was shown to play a relevant role with respect to the 
whole evacuation delays \cite{Dorso1,Dorso2}. However, either the amount of 
blocking structures or its time life can vary according to the 
door width, the presence of obstacles or fallen individuals 
\cite{Dorso3,Cornes1}. Further studies on blocking structures appearing in 
granular media research can be found in 
Refs.~\cite{Garcimartin1,Garcimartin2,Garcimartin3,Garcimartin4}.  \\

The relevance of the \textit{blocking structures} on the time evacuation 
performance has alerted researchers that the analysis of ``reduced'' 
systems rather than the whole crowd is still a meaningful approach to the FIS 
effect. In this context, the authors of Ref.~\cite{suzuno} introduced a 
simplified breakup model for a small arch-like blocking structure (in a SFM 
setting). They examined theoretically the breakup of the arch due to a single 
moving particle, and observed a FIS-like behavior. Thus, they concluded that 
the essentials of the FIS phenomenon could be described with a system of only a 
few degrees of freedom.  \\

To our knowledge, neither the theoretical approach nor the computational 
simulations have been pushed to extreme scenarios. That is, no special 
attention has been paid to those situations where the pedestrians experience
very high anxiety levels (see Section~\ref{results}) while the crowd 
becomes increasingly large. \\

In the current investigation we explore the pedestrians anxiety levels from a 
relaxed situation to desired velocities that may cause dangerous pressures. A 
dangerous pressure of 1600 Nm$^{-1}$ may be associated to at least three 
pedestrians pushing with a desired velocity close to 20~m/s (see 
Refs.~\cite{Helbing1,nasa1995}). \\

Our work is organized as follows: a brief review of the basic SFM can be found 
in Section~\ref{background}. Section~\ref{simulations} details the simulation 
procedures used to studying the room evacuation of a crowd under panic. The 
corresponding results are presented in Section~\ref{results}. Finally, the 
conclusions are summarized in Section~\ref{conclusions}. \\

\section{\label{background}Background}

\subsection{\label{sfm}The Social Force Model}

Our research was carried out in the context of the ``social force model'' (SFM) 
proposed by Helbing and co-workers \cite{Helbing1}. This model states that human 
motion is caused by the desire of people to reach a certain destination, as 
well as other environmental factors. The pedestrians behavioral pattern in a 
crowded environment can be modeled by three kind of forces: the ``desire 
force'', the ``social force'' and the ``granular force''. \\

The ``desire force'' represents the pedestrian's own desire to reach a 
specific target position at a desired velocity $v_d$. But, in order to reach 
the desired target, he (she) needs to accelerate (decelerate) from his (her) 
current velocity $\mathbf{v}^{(i)}(t)$. This acceleration (or deceleration) 
represents a ``desire force'' since it is motivated by his (her) own 
willingness. The corresponding expression for this forces is 

\begin{equation}
        \mathbf{f}_d^ {(i)}(t) =  
m_i\,\displaystyle\frac{v_d^{(i)}\,\mathbf{e}_d^
{(i)}(t)-\mathbf{v}^{(i)}(t)}{\tau} \label{desired}
\end{equation}

\noindent where $m_i$ is the mass of the pedestrian $i$. $\mathbf{e}_d$ 
corresponds to the unit vector pointing to the target position and $\tau$ is a 
constant related to the relaxation time needed to reach his (her) desired 
velocity. Its value is determined experimentally. For simplicity, we assume that 
$v_d$ remains constant during an evacuation process and is the same for all 
individuals, but $\mathbf{e}_d$ changes according to the current position of the 
pedestrian. Detailed values for $m_i$ and $\tau$ can be found in 
Refs.~\cite{Helbing1,Dorso3}. \\

The ``social force'' represents the psychological tendency of two pedestrians,  
say $i$ and $j$, to stay away from each other by a repulsive interaction force 

\begin{equation}
        \mathbf{f}_s^{(ij)} = A_i\,e^{(r_{ij}-d_{ij})/B_i}\mathbf{n}_{ij} 
        \label{social}
\end{equation}

\noindent where $(ij)$ means any pedestrian-pedestrian pair, or pedestrian-wall 
pair. $A_i$ and $B_i$ are fixed values, $d_{ij}$ is the distance between  the 
center of mass of the pedestrians $i$ and $j$ and the distance $r_{ij}=r_i+r_j$ 
is the sum of the pedestrians radius. $\mathbf{n}_{ij}$ means the unit vector in 
the $\vec{ji}$ direction. \\

Any two pedestrians touch each other if their distance $d_{ij}$ is smaller than 
$r_{ij}$.  In this case, an additional force is included in the model, called 
the ``granular force''. This force is considered be a linear function of the 
relative (tangential) velocities of the contacting individuals. Its mathematical 
expression reads 

\begin{equation}
        \mathbf{f}_g^{(ij)} = 
\kappa\,(r_{ij}-d_{ij})\,\Theta(r_{ij}-d_{ij})\,\Delta
\mathbf{v}^{(ij)}\cdot\mathbf{t}_{ij} 
        \label{granular}
\end{equation}

\noindent where $\kappa$ is a fixed parameter. The function 
$\Theta(r_{ij}-d_{ij})$ is zero when its argument is negative (that is, 
$r_{ij}<d_{ij}$) and equals unity for any other case (Heaviside function). 
$\Delta\mathbf{v}^{(ij)}\cdot\mathbf{t}_{ij}$ represents the difference between 
the tangential velocities of the sliding bodies (or between the individual and 
the walls).   \\

The above forces actuate on the pedestrians dynamics by changing his (her) 
current velocity. The equation of motion for pedestrian $i$ reads

\begin{equation}
m_i\,\displaystyle\frac{d\mathbf{v}^{(i)}}{dt}=\mathbf{f}_d^{(i)}
+\displaystyle\sum_{j=1}^{N}\displaystyle\mathbf{f}_s^{(ij)}
+\displaystyle\sum_ {
j=1}^{N}\mathbf{f}_g^{(ij)}\label{eq_mov}
\end{equation}

\noindent where the subscript $j$ represents all the other pedestrians 
(excluding $i$) and the walls. \\

\subsection{\label{human}Clustering structures}

The time delays during an evacuation process are related to clogged people, 
as explained in Refs.~\cite{Dorso1,Dorso2}. Groups of pedestrians can 
be defined as the set of individuals that for any member of the group (say, 
$i$) there exist at least another member belonging to the same group ($j$) 
in contact with the former. That is, the distance between them ($d_{ij}$) is 
less than the sum of their radius ($d_{ij}<r_i+r_j$). This kind of structure is 
called a \textit{human cluster} and it can be mathematically defined as  

\begin{equation}
 i\in\mathcal{G} \Leftrightarrow \exists j\in\mathcal{G}/d_{ij}<r_i+r_j
\end{equation}

\noindent where $\mathcal{G}$ corresponds to any set of individuals. \\

\begin{figure}
\includegraphics[scale=0.3]{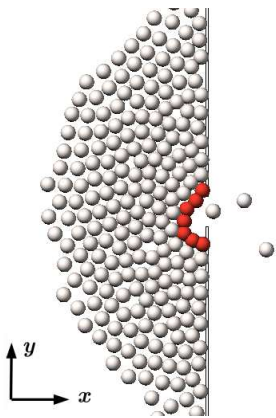}
\caption{\label{fig:5} (Color on-line only) Snapshot of an evacuation process 
from a $20\,\mathrm{m}\times20\,$m room, with a single door of 1.2~m width. The 
blocking structure is identified in red color. The rest of the crowd is 
represented by white circles. It can be seen three individuals that have 
already left the room. The desired velocity for the individuals inside the 
room was $v_d=6\,$m/s.  }
\end{figure}

During an evacuation process, different human clusters may appear inside the 
room. But, some of them are able to \textit{block} the way out. We are 
interested in the minimum set of human clusters that connects both sides of the 
exit. Thus, we will call \textit{blocking clusters} or \textit{blocking 
structures} to those human structures that block the exit. Two blocking clusters 
are different if they differs at least in one pedestrian. That is, if they 
differ in the number of members or in pedestrians themselves. Fig.~\ref{fig:5} 
shows (in highlighted color) a \textit{blocking structure} near the door. \\

We define the \textit{blocking time} as the total time during which the 
evacuation process is stopped due to any blocking cluster. That is, the sum of 
the ``life time'' of each blocking cluster (\textit{blocking delays}). \\

\section{\label{simulations}Numerical simulations}

Most of the simulation processes were performed on a 20~m $\times$ 20~m room with 
225 pedestrians inside. The occupancy density was close to 
0.6~individuals/m$^2$ as suggested by healthy indoor environmental regulations 
\cite{mysen}. The room had a single exit on one side, as shown in 
Fig.~\ref{fig:5}. The door was placed in the middle of the side wall to 
avoid corner effects.\\

A few simulation processes were performed on 30~m $\times$ 30~m and 40~m 
$\times$ 40~m rooms with 529 and 961 pedestrians inside, respectively. The door 
was also placed in the middle of the side wall. 
\\

The pedestrians were initially placed in a regular square arrangement  
along the room with random velocities, resembling a Gaussian distribution 
with null mean value. The desired velocity $v_d$ was the same for all the 
individuals. At each time-step, however, the desired direction $\mathbf{e}_d$ 
was updated, in order to point to the exit. \\

Two different boundary conditions were examined. The first one included a 
re-entering mechanism for the outgoing pedestrians. That is, those individuals 
who were able to leave the room were moved back inside the room and placed at  
the very back of the bulk with velocity $v=0.1\,$m/s, in order to cause a 
minimal bulk perturbation. This mechanism was carried out in order to keep 
the crowd size unchanged. \\

The second boundary condition was the open one. That is, the individuals 
who left the room were not allowed to enter again. This condition approaches to 
real situations, and thus, it is useful for comparison purposes.\\

The simulating process lasted for approximately 2000~s whenever the re-entering 
mechanism was implemented. If no re-entering was allowed, each evacuation 
process lasted until 70\% of individuals had left the room. If this condition 
could not be fulfilled, the process was stopped after 1000~s. Whenever the 
re-entering  mechanism was not allowed, at least 30 evacuation processes were 
run for each desired velocity $v_d$. \\ 

The explored anxiety levels ranged from relaxed situations ($v_d<2\,$m/s) to 
extremely  stressing ones ($v_d=20\,$m/s). This upper limit may hardly be 
reached in real life situations. However, extremely stressing situations may 
produce similar pushing pressures as those in a larger crowd with moderate 
anxiety levels (see Ref.~\cite{Dorso5} for details). Thus, this wide range of 
desired velocities provided us a full picture of the blocking effects due to 
high pressures. \\

The simulations were supported by {\sc Lammps} molecular dynamics simulator 
with parallel computing capabilities \cite{plimpton}. The time integration 
algorithm followed the velocity Verlet scheme with a time step of $10^{-4}\,$s. 
All the necessary parameters were set to the same values as in previous works 
(see Refs.~\cite{Dorso3,Dorso4,Dorso5}). \\

We implemented special modules in C++ for upgrading the {\sc Lammps} 
capabilities to attain the ``social force model'' simulations. We also checked 
over the {\sc Lammps} output with previous computations (see 
Refs.~\cite{Dorso3,Dorso4}).\\

Data recording was done at time intervals of $0.05~\tau$, that is, at intervals 
as short as 10\% of the pedestrian's relaxation time (see Section 
\ref{sfm}). The recorded magnitudes were the pedestrian's positions and 
velocities for each evacuation process. We also recorded the corresponding 
social force $f_s$ and granular force $f_g$ actuating on each individual. \\

\section{\label{results}Results}

\subsection{\label{evacuation}Evacuation time versus the desired velocity}

\begin{figure}
\includegraphics[width=\columnwidth]
{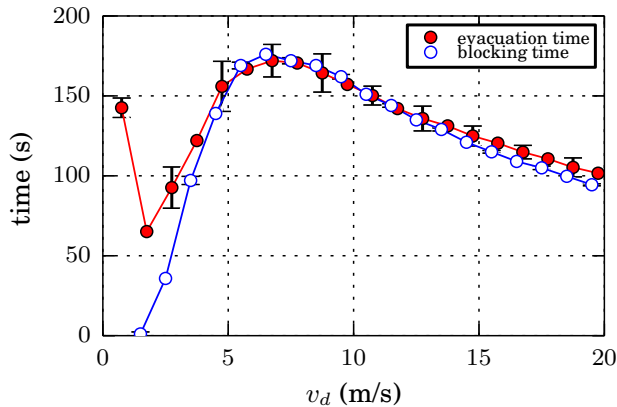}
\caption{\label{fig:11} (Color on-line only) Evacuation time and blocking 
time as a function of the desired velocity $v_d$. Both data sets represent the 
mean values from 60 evacuation processes. The simulated room was 20$\times$20~m 
with a single door of 1.2~m width on one side. The number of individuals inside 
the room was 225 (no re-entering mechanism was allowed). The simulation lasted 
until 160 individuals left the room.  }
\end{figure}

As a first step we measured the mean evacuation time for a wide range of  
desired velocities $v_d$, many of them beyond the interval analyzed by Helbing 
and co-workers (see Ref.~\cite{Helbing1}). This is shown in Fig.~\ref{fig:11} 
(filled symbols and red line). The \textit{faster is slower} regime can be 
observed for desired velocities between 2~m/s and 8~m/s (approximately). 
However, the evacuation time improves beyond this interval, meaning that the 
greater the pedestrian's anxiety level, the better with respect to 
the overall time saving. This phenomenon was reported for both boundary 
conditions mentioned in Section~\ref{simulations}. \\

Therefore, we actually attain a \textit{faster is faster} regime for desired 
velocities larger than 8~m/s, instead of the expected \textit{faster is 
slower} regime. This is a novel behavior that has not been reported before (to 
our knowledge) in the literature. This effect holds even if we include the 
elastic force introduced by Helbing et al. in Ref.~\cite{Helbing1} (not shown 
in Fig.~\ref{fig:11}). \\ 

The overall time performance has been reported to be related to the 
\textit{clogging delays}, understood as the period of time between two outgoing 
pedestrians (see Refs.~\cite{Dorso1,Dorso2,Dorso3} for details). But, since 
most of these time intervals correspond to the presence of \textit{blocking 
structures} near the door, we examined closely the delays due to blockings 
for increasing anxiety levels (\textit{i.e.} desired velocities $v_d$).  \\

Fig.~\ref{fig:11} exhibits (in hollow symbols and blue line) the computed 
blocking time for a wide range of desired velocities. That is, the 
cumulative ``life time'' of all the \textit{blocking clusters} occurring 
during an evacuation process. Notice that the blocking delays become 
non-vanishing for $v_d>2\,$m/s. This threshold corresponds to those situations 
where the granular forces become relevant, according to 
Refs.~\cite{Dorso1,Dorso2}. It is, indeed, the lower threshold for the 
\textit{faster is slower} effect. \\

No complete matching between the mean evacuation time and the blocking time can 
be observed along the interval $2\,\mathrm{m/s}<v_d<4\,\mathrm{m/s}$. This 
means that the blocking time does not fulfill the evacuation time, but other 
time waists are supposed to be relevant. We traced back all the time delays 
experienced by the pedestrian, and noticed that the time lapse between the 
breakup of the blocking structure and the leaving time of the pedestrians 
(belonging to this blocking structure) was actually a relevant magnitude. This 
\textit{transit time} explained the difference between the evacuation time and 
the blocking time. \\

According to Fig.~\ref{fig:11}, the transit time does not play a role for
desired velocities larger than $v_d=4\,$m/s. The evacuation time appears to be 
highly correlated to the blocking delays above this value. Thus, the noticeable 
enhancement in the evacuation performance taking place between $8\,$m/s and 
$20\,$m/s (\textit{i.e.} the ``faster is faster'' effect) is somehow related 
to the enhancement in the blocking time. In other words, the delays associated 
to the blocking clusters appear to explain the entire \textit{faster is 
faster} effect.  \\

\begin{figure}
\includegraphics[width=\columnwidth]
{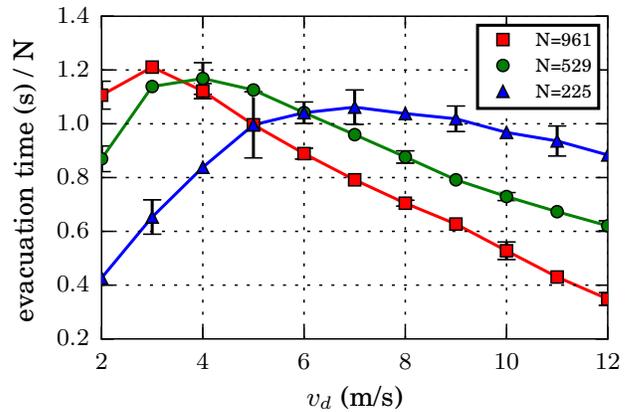}
\caption{\label{fisf_many} (Color on-line only) Evacuation time per individual 
vs. desired velocity for N$=$225, N$=$529 and N$=$961 (no re-entering mechanism 
was allowed). The rooms sizes were 20$\times$20~m, 
30$\times$30~m and 40$\times$40~m, respectively, with a single door of 1.2~m 
width on one side. Mean values were computed from 30 evacuation processes until 
70\% of pedestrians left the room.  }
\end{figure}

We next measured the evacuation time for three different crowd sizes. We chose 
a relatively small crowd (225 pedestrians), a moderate one (529 pedestrians) 
and a large one (961 pedestrians). The corresponding room sizes were 
20$\times$20~m, 30$\times$30~m and 40$\times$40~m, respectively. The results are 
shown in Fig.~\ref{fisf_many}. \\

The three situations exhibited in Fig.~\ref{fisf_many} achieve a \textit{faster 
is faster} phenomenon, since the slope of each evacuation curve changes sign 
above a certain desired velocity. As the number of individuals in the crowd 
becomes larger, the $v_d$ interval attaining a negative slope increases. That 
is, only a moderate anxiety level is required to achieve the \textit{faster 
is faster} phenomenon if the crowd is large enough. \\

Notice that the larger crowd (\textit{i.e.} 961 individuals) attains the 
steepest negative slope. Thus, as more people push to get out (for any fixed 
desired velocity $v_d$), the faster they will evacuate. \\

\begin{figure}
\includegraphics[width=\columnwidth]
{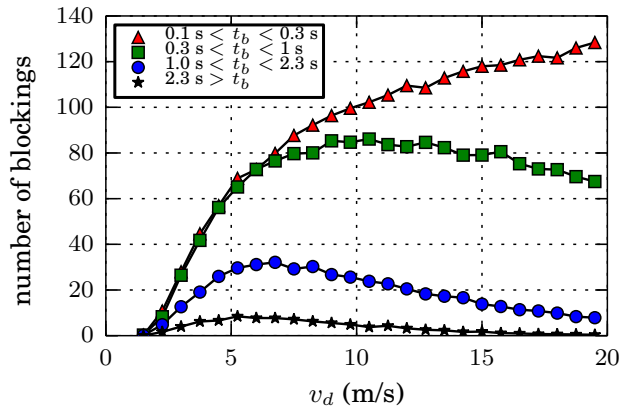}
\caption{\label{blocking_bin} (Color on-line only) Mean number of blocking 
delays for four different time intervals (see legend for the 
corresponding blocking times $t_b$) as a function of the desired velocity 
$v_d$. The simulated room was 20$\times$20~m with a single door of 1.2~m width 
on one side. The number of individuals inside the room was 225 (no re-entering 
mechanism was allowed). Mean values were computed from 60 realizations. The 
simulation lasted until 160 individuals left the room.}
\end{figure}

For a better insight of the ``faster is faster'' phenomenon, we binned the 
blocking delays into four time intervals or categories. This allowed a 
quantitative examination of the changes in the delays when moving from the 
``faster is slower'' regime to the ``faster is faster'' regime. 
Fig.~\ref{blocking_bin} shows the mean number of  blocking delays (for each 
time interval) as a function of $v_d$. \\

The four blocking time intervals represented in Fig.~\ref{blocking_bin} 
increase for increasing desired velocities until $8\,$m/s. This is 
in agreement with the ``faster is slower'' regime, since the faster the 
pedestrians try to evacuate, the more time they spend stuck in the blocking 
structure. \\

Beyond 8~m/s, the number of blockings corresponding to those time intervals 
greater than 0.3$\,$s reduces (as $v_d$ increases). Thus, the individuals spend 
less time stuck in the blocking structure for increasing anxiety levels. \\

It is true that the delays between 0.1$\,$s and 0.3$\,$s increase for high 
anxiety levels. But a quick inspection of Fig.~\ref{blocking_bin} shows that 
this increase (represented in red triangular symbols) is not enough to balance 
the decrease in the time intervals greater than 0.3$\,$s. Consequently, the 
overall evacuation time follows the same behavior as the long lasting delays 
(say, the \textit{faster is faster} behavior).  \\ 

The above research may be summarized as follows. The scenario for high 
anxiety levels (say, $v_d>4\,$m/s) corresponds to a ``nearly always'' 
blocking scenario. However, two different blocking instances can be noticed. The 
``faster is slower'' corresponds to the first instance. The ``faster is faster'' 
is the second instance appearing after either high values of $v_d$ or increasing 
number of pedestrians. Many long lasting blockings seem to break down into 
shorter blockings, or even disappear (see Fig.~\ref{blocking_bin}).  \\

Our results, so far, suggest that the breakup process of the blocking structures 
needs to be revisited.  We hypothesize that a connection between this breakup 
process and the pedestrian's pushing efforts should exist. The next two 
Sections will focus on this issue. \\

\subsection{\label{breakup}Blocking cluster breakup}

\begin{figure}
\includegraphics[width=\columnwidth]
{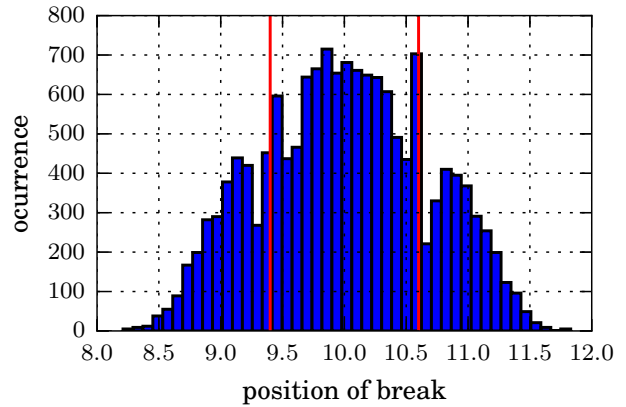}
\caption{\label{histo} (Color on-line only) Histogram of the position of the 
breakup of the blocking cluster. The room size was 20$\times$20~m with 225 
pedestrians (no re-entering mechanism was allowed). The door's width was 1.2~m 
(from $y=9.4$~m to $y=10.6$~m). The vertical red lines represent it's limits. 30 
evacuation processes were performed until 70\% of pedestrians left the room. The 
desired velocity was $v_d=$10~m/s }
\end{figure}

We now examine the position of the breakups in the blocking cluster. We define 
the \textit{breakup position} as the one on the $y$-axis (according to 
Fig.~\ref{fig:5}) where any pedestrian gets released from the blocking 
structure. Fig.~\ref{histo} exhibits a histogram of the breakup position for a 
fixed anxiety level ($v_d=10\,$m/s).  \\

The mean value of the distribution in Fig.~\ref{histo} is close to $y=10\,$m, 
that is, the mid-position of the door. This means that the breakups are 
likely to occur in front of the exit. The same result holds for other 
desired velocities in the investigated range (not shown). Therefore, this 
region is of special interest with respect to the breakup process.\\

From our current simulations and previous work (see Ref.~\cite{Dorso5}), we 
realized that the mid-position corresponds to the crowd area of highest 
pressure (for an exit width of 1.2~m). This is in agreement with the maximum 
amount of breakups, since higher pushing efforts may help forward the blocking 
pedestrians. \\

\subsection{\label{stationary}Stationary blocking model}

For a better understanding of the relation between the crowd pushing forces 
and the breakup process, we decided to focus on the behavior of a single 
pedestrian who tries to get released from the blocking structure. \\

We mimicked a small piece of the blocking structure (\textit{i.e.} red 
individuals in Fig.~\ref{fig:5}) as two individuals standing still, 
but separated a distance smaller than the pedestrian's diameter. A third 
pedestrian was set in between the former, mimicking the pedestrian who tries to 
get released from the blocking structure. Fig.~\ref{fig:1} represents this set 
of three pedestrians. Notice that Fig.~\ref{fig:1} may represent any piece of 
the blocking structure, but according to Section~\ref{breakup}, it will usually
correspond to the middle piece of the blocking structure. \\

The middle pedestrian in Fig.~\ref{fig:1} is being pushed from behind by the 
rest of the crowd. The crowd pushing force $f_s$ points in the 
$x$-direction. Two granular forces $f_g$ appear in the opposite direction as 
a consequence of pedestrian's advancement. More details can be found in 
Appendix~\ref{appendix}. \\  

The still pedestrians on both sides experience the repulsion due to the 
mid-pedestrian, as shown in Fig.~\ref{fig:2}. This repulsion $f$ points in the 
$y$-direction. We are assuming, however, that the pedestrians on the sides do 
not move during the breakup process. Thus, the force $f$ should be balanced by 
the crowd (in the $y$-direction). This corresponds to the balancing force 
$\mathcal{F}$ in Fig.~\ref{fig:2}. More details can be found in 
Appendix~\ref{appendix}. \\

Notice that our mimicking model assumes that the crowd pushes the mid-pedestrian 
along the $x$-direction, while also pushes the still pedestrians along the 
$y$-direction. Both forces ($f_s$ and $\mathcal{F}$) are similar in nature. 
Actually, for the current geometry, $f_s$ and $\mathcal{F}$ are approximately 
equal. \\

The crowd pushing force increases for increasing anxiety levels. For a slowly 
moving crowd, this force varies linearly with $v_d$, according to 
Eq.~(\ref{desired}). We can therefore set its value as

\begin{equation}
 f_s=\mathcal{F}=\beta v_d\label{crowd_value}
\end{equation}

\noindent for any fixed coefficient $\beta$. The value of $\beta$ depends 
linearly on the number of individuals in the crowd. \\

We assume a completely blocked situation at the beginning of the simulation. 
The center of mass of the three pedestrians were initially aligned and the 
velocity of the individual in the middle was set to zero. \\

\begin{figure}
\includegraphics[width=\columnwidth]
{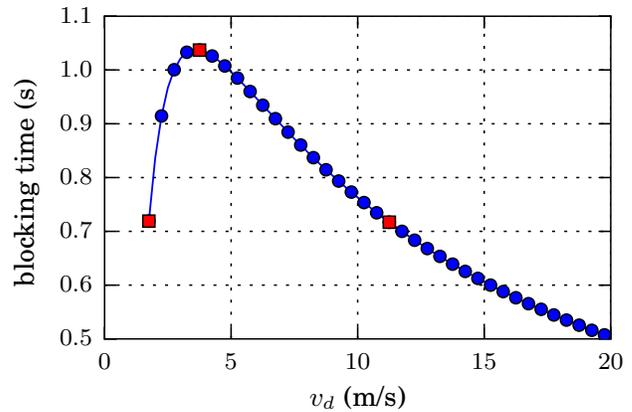}
\caption{\label{fisf_narrow_nuevo} (Color on-line only) Blocking time of the 
three pedestrians model (one moving pedestrian between two still ones) as a 
function of the desired velocity $v_d$. The initial velocity of the moving 
pedestrian was set to zero. The crowd pressure was set to 
$\mathcal{F}=f_s=2000\,v_d$. Each blocking time was recorded when the 
moving pedestrian lost contact with the other individuals. Desired velocities 
of $v_d=1.75$~m/s, $v_d=3.5$~m/s and $v_d=11.25$~m/s are indicated in red 
color (and squared symbols). The blocking time for $v_d=1.75$~m/s and 
$v_d=11.25$~m/s are the same. Only one realization was done for each $v_d$ 
value.}
\end{figure}

We computed the blocking time on this simple model. This was defined as the 
period of time required for the moving pedestrian to release from the 
other two (still ones). This time is supposed to mimic the blocking time of the 
blocking structure, since the three pedestrians represent a small piece of this 
 structure. Fig.~\ref{fisf_narrow_nuevo} shows the blocking time as a function 
of the desired velocity $v_d$. \\

A comparison between Fig.~\ref{fig:11} and Fig.~\ref{fisf_narrow_nuevo} shows 
the same qualitative behavior for the blocking time, although the scale along 
the $v_d$ axis is somehow different. The blocking time slope changes sign at 
7~m/s in  Fig.~\ref{fig:11}, while Fig.~\ref{fisf_narrow_nuevo} shows a 
similar change at 3.75~m/s. This discrepancy can be explained because of the 
chosen value of $\beta$. \\

The chosen value for $\beta$ in Fig.~\ref{fisf_narrow_nuevo} was 2000 (see 
caption). This value corresponds to the expected pushing force for a crowd of 
225 pedestrians (and $v_d=2\,$m/s). However, as the pedestrians evacuate from 
the room, the crowd pushing force diminishes. The effective force along the 
whole process is actually smaller, and so is the $\beta$ value. Thus, according 
to Eq.~(\ref{eq:a7}), the ``effective'' maximum blocking time is expected to 
lie at a larger $v_d$ value than 3.75~m/s.    \\

The above reasoning is also in agreement with the evacuation time shown in 
Fig.~\ref{fisf_many} for an increasing number of pedestrians. The maximum 
evacuation time takes place at lower anxiety levels (\textit{i.e.} $v_d$ 
values) as the crowd size becomes larger. Therefore, the  pushing force $\beta 
v_d$  downscales the \textit{faster is faster} threshold, as expected from our 
simple model. \\

So far, the mimicking model for a small piece of the blocking structure 
exhibits a \textit{faster is slower} instance for low crowd's pushing forces, 
and a \textit{faster is faster} instance for large pushing forces. The 
associated equations for both instances are summarized in 
Appendix~\ref{appendix}. This formalism, however, stands for a simple 
stationary situation. We will release this hypothesis in Section 
\ref{non_stationary}. \\

\subsection{\label{non_stationary}Non-stationary blocking model}

We were able to establish a connection between the breakup process and the 
crowd pushing forces in Section \ref{stationary}. Now, we will examine the 
force balance on the moving pedestrian along the $x$-axis (see 
Fig.~\ref{fig:1}). As already mentioned, our attention is placed on initially 
aligned pedestrians with null velocity.  \\

\begin{figure}
\includegraphics[width=\columnwidth]
{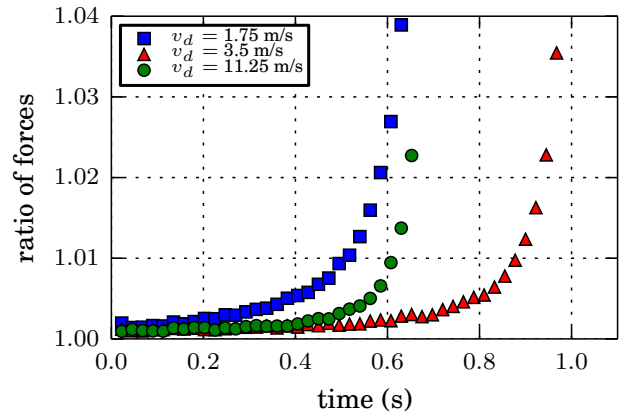}
\caption{\label{cociente_fuerzas_nuevo} (Color on-line only) Ratio of 
\textit{positive forces} (desire force and social repulsion) and 
\textit{negative force} (granular) on the moving pedestrian as a function of 
time for three desired velocities (see text for details). The initially 
velocity of the moving pedestrian was zero. The simulation finished when he 
looses contact with the other individuals. One realization is done.}
\end{figure}

Fig.~\ref{cociente_fuerzas_nuevo} shows the force balance on the moving 
pedestrian (of the mimicking model) during the simulated breakup process. The 
balance is expressed as the ratio between the \textit{positive forces} and the 
\textit{negative forces}. The former corresponds to the sum of all the forces 
that push the moving pedestrian towards the exit (\textit{i.e.} the own desired 
force, and the social force from all the neighbors). The latter corresponds to 
the force in the opposite direction to the movement (\textit{i.e} the granular 
force). According to Section~\ref{sfm} and Fig.~\ref{fig:1and2} \\

\begin{equation}
 \mathrm{ratio}=\displaystyle\frac{f_s+\mathcal{F}+f_d}{2f_g}\label{ratio}
\end{equation}

\noindent where $f_s$ and $\mathcal{F}$ correspond to the pushing forces 
from the crowd. Both are 
social forces in nature. Notice, however, that only the contribution on the 
$x$-axis is relevant in the mimicking model (see Fig.~\ref{fig:1and2}). \\

Fig.~\ref{cociente_fuerzas_nuevo} presents three different situations, 
corresponding  to those desired velocities highlighted in red color in 
Fig.~\ref{fisf_narrow_nuevo}. The three situations stand for any \textit{faster 
is slower} instance, the maximum blocking time instance and any \textit{faster 
is faster} instance, respectively. But care was taken in choosing similar 
blocking times for the first and the third situation, in order to achieve a fair 
comparison.  \\

The three situations shown in Fig.~\ref{cociente_fuerzas_nuevo} exhibit a ratio 
close to unity during the first stage of the process. This means that all the 
forces actuating on the moving pedestrian are approximately balanced. The 
formalism presented in Appendix~\ref{appendix} is approximately valid during 
this stage of the process. \\

Notice that this quasi-stationary stage lasts until the very end 
of the breakup process (say, 1\% above unity).  However, a striking 
positive slope can be seen during the last stage of each process. The 
slopes are quite similar on each process (although shifted in time), and thus, 
 this last stage seems not to be relevant in the overall blocking time. We can 
envisage the last stage as an expelling process before the blocking structure 
breaks into two pieces.    \\

An important conclusion can be derived from the inspection of 
Fig.~\ref{cociente_fuerzas_nuevo}: although the breakup process is actually 
non-stationary, the balance constrain (ratio$\,\simeq\,1$) is quite accurate 
for the early breakup process. \\

\subsection{\label{remarks}Remarks}

From our point of view, the balance constrain (that is, ratio$\,\simeq\,$1) is 
actually the main reason for the \textit{faster is faster} phenomenon to take 
place. \\

Recall that the \textit{positive} forces $f_s+\mathcal{F}+f_d$ correspond to 
the sum of the pushing forces of the crowd ($f_s$ and $\mathcal{F}$) and the 
moving pedestrian's own desire ($f_d$). The latter, however, is not relevant 
with respect to the former because most of the pushing effort is done by the 
crowd (for example, $f_d$ is approximately 10\% of $f_s$ for  225 individuals). 
Thus, the \textit{positive} forces are roughly $f_s+\mathcal{F}=2\beta v_d$, 
according to Section~\ref{stationary} and Appendix~\ref{context}. The balance 
constrain becomes approximately

\begin{equation}
\displaystyle\frac{\beta v_d}{f_g}\simeq 1\label{constrain}
\end{equation}

Eq.~(\ref{constrain}) is meaningful since it expresses the fact that the 
\textit{negative} force $f_g$  balances the pushing force, in order to 
keep the pedestrian moving forward (at an almost constant velocity). 
However, the granular force is currently $f_g=\kappa\,v\,B\ln(\beta v_d/A)$. 
The $B\ln(\beta v_d/A)$ factor corresponds to the compression between the 
pedestrian and his (her) neighbor in the blocking structure (see 
Eq.~(\ref{eq:a5}) for details). Thus    \\

\begin{equation}
v^{-1}\sim \displaystyle\frac{\ln(\beta 
v_d/A)}{\beta v_d/A}\label{constrain2}
\end{equation}

Notice that Eq.~(\ref{constrain2}) resembles the behavior of 
Fig.~\ref{fisf_narrow_nuevo}. The slope of $v^{-1}$ is positive for low 
anxiety levels (\textit{i.e.} $v_d$ values), but changes sign as the anxiety 
level becomes increasingly large. Since the blocking time varies as $v^{-1}$, 
we may conclude that Eq.~(\ref{constrain2}) mimics the \textit{faster is 
slower} and the \textit{faster is faster} instances. \\

The logarithm in Eq.~(\ref{constrain2}) is the key feature for the slope 
change. Recall from Eq.~(\ref{eq:a5}) that $\ln(\beta v_d/A)$ stands for the 
compression in the blocking structure. But, although compression increases for 
increasing pushing forces of the crowd, it seems not enough to diminish the 
pedestrian velocity in order to hold the \textit{faster is slower} phenomenon 
at high anxiety levels. Consequently, the blocking time decreases, achieving a 
\textit{faster is faster} instance. \\   

In Section~\ref{appendix_remaks} a more detailed formalism is exhibited on this 
issue. \\


\section{\label{conclusions}Conclusions}

Our investigation focused on the evacuation of extremely anxious 
pedestrians through a single emergency door, in the context of the ``social 
force model''. No previous research has been done, to our knowledge, for 
anxiety levels so high that may cause dangerous pressures (even in relatively 
small crowds).\\

Unexpectedly, we found an improvement in the overall evacuation time for 
desired velocities above $8\,$m/s (and a crowd size of 225 individuals). 
That is, the \textit{faster is slower} effect came to an end at this anxiety 
level, while a novel \textit{faster is faster} phenomenon raised (at least) 
until a desired velocity of $20\,$m/s. This unforeseen phenomenon was also 
achieved for increasingly large crowds and lower desired velocities. \\

A detailed examination of the pedestrian's blocking clusters showed that the 
\textit{faster is faster} instance is related to shorter ``life times'' of 
the blocking structures near the exit. The long lasting structures 
taking place at the \textit{faster is slower} instance now breakup into short 
lasting ones. The breakup is most likely to occur straight in front of the 
exit.  \\ 

We mimicked the breakup process of a small piece of the blocking structure 
through a minimalistic model. The most simple model that we could image was a 
moving pedestrian between two still individuals. Although its simplicity, it 
was found to be useful for understanding the connection between the 
crowd's pushing forces and the blocking breakup process.\\

The mimicking model for the blocking structure showed that a balance between the 
crowd's pushing forces and the friction with respect to the neighboring 
individuals held along the breakup. Only at the very end of the process, the 
pedestrian was expelled out of the blocking structure. \\

We concluded from the force balance condition that friction was the
key feature for the \textit{faster is faster} instance to take place. As the 
crowd pushing force increases, the compression between individuals in the 
blocking structure seems not enough to provide a slowing down in the moving 
pedestrian. Thus, the \textit{faster is slower} instance switches to a 
\textit{faster is faster} instance. The latter can be envisaged as brake 
failure mechanism. \\

We want to stress the fact that, although we investigated extremely high 
anxiety situations, \textit{faster is faster} instance may be present at 
lower desired velocities if the crowd size is large enough. We were able to
acknowledge the \textit{faster is faster} phenomenon for desired 
velocities as low as $v_d=4\,$m/s when the crowd included 1000 individuals 
approximately. \\

\begin{acknowledgments}
This work was supported by the National Scientific and Technical 
Research Council (spanish: Consejo Nacional de Investigaciones Cient\'\i ficas 
y T\'ecnicas - CONICET, Argentina) grant number PIP 2015-2017 GI, founding 
D4247(12-22-2016). C.O.~Dorso is full researcher of the CONICET. G.A.~Frank is 
 assistant researcher of the CONICET. I.M.~Sticco and F.E.~Cornes 
have degree in Physics.
\end{acknowledgments}

\appendix

\section{\label{appendix}A simple blocking model}

\subsection{The dynamic}

\begin{figure*}[!htbp]
\subfloat[force balance for the $x$-axis.\label{fig:1}]{
\includegraphics[width=0.85\columnwidth]{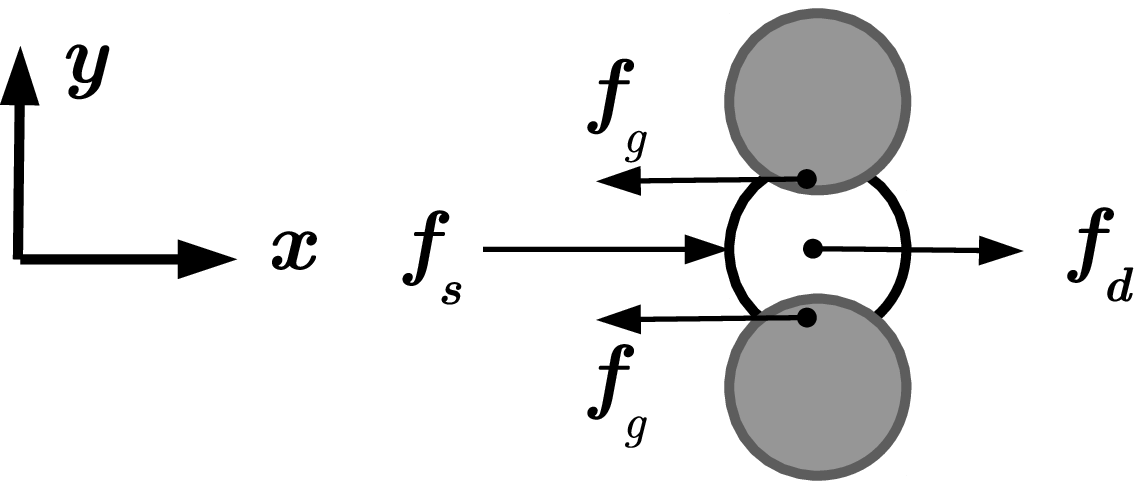}
}\hfill
\subfloat[force balance for the $y$-axis. \label{fig:2}]{
\includegraphics[width=0.85\columnwidth]{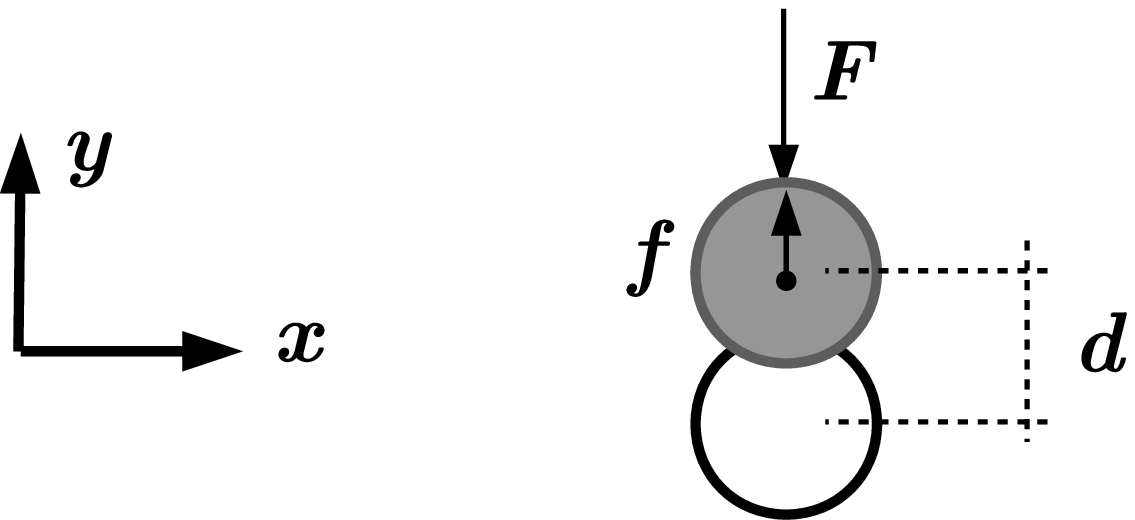}
}
\caption{\label{fig:1and2} Force balance for a moving pedestrian between two 
still individuals. The moving pedestrian is represented by the white circle, 
while the gray circles correspond to the still individuals. The movement is 
in the $+x$ direction. $f_s$ represents the (mean) force due to other 
pedestrians pushing from behind. $f_d$ is the moving pedestrian's own desire. 
$f_g$ corresponds to the tangential friction (\textit{i.e.} granular force) 
between the moving pedestrian and his (her) neighbors. $\mathcal{F}$ and $f$ are 
the forces actuating on the upper (still) pedestrian. $f$ corresponds to the 
social repulsive force due to the moving pedestrian, while $\mathcal{F}$ 
represents the counter force for keeping the pedestrian still. } 
\end{figure*}

This Appendix examines in detail a very simple model for the time delays in the 
blocking cluster. We consider a single moving pedestrian stuck in the blocking 
cluster, as shown in Fig.~\ref{fig:1and2}. The moving pedestrian tries to get
released from two neighboring individuals that are supposed to remain still 
during the process. The three pedestrians belong to the same blocking 
structure, according to the definition given in Section~\ref{human}. The 
equation of motion for the pedestrian in the middle of Fig.~\ref{fig:1} reads

\begin{equation}
m\,\displaystyle\frac{dv}{dt}=f_s+f_d-2f_g\label{eq:a1}
\end{equation}

\noindent where $f_s$ represents the force due to other pedestrians pushing 
from behind, $f_d$ represents the moving pedestrian own desire, and $f_g$ 
represents the corresponding tangential friction due to contact between the 
neighboring pedestrians. $m$ and $v$ are the mass and velocity of the moving 
pedestrian (see caption in Fig.~\ref{fig:1and2}), respectively. The 
expressions for $f_d$ and $f_g$ are as follows\\

\begin{equation}
\left\{\begin{array}{lcl}
        f_d=\displaystyle\frac{m}{\tau}(v_d-v) \\
        & & \\
        f_g=\kappa\,(2r-d)\,v & \mathrm{if} & 2r-d>0 \\
       \end{array}\right.\label{eq:a2}
\end{equation}

The granular force $f_g$ expressed in (\ref{eq:a2}) depends only on the 
velocity $v$ since the other pedestrians are supposed to remain still. The 
magnitude $2r-d$ is the difference between the pedestrian's diameter $2r$ and 
the inter-pedestrian distance $d$. It represents the compression between two 
contacting individuals. The other parameters correspond to usual literature 
values (see Refs.~\cite{Dorso1,Dorso2}).\\ 

The movement equation (\ref{eq:a1}) expresses the dynamic for the passing 
through pedestrian. The characteristic time needed for the pedestrian to reach 
the stationary state is

\begin{equation} 
t_c=\displaystyle\frac{\tau}{1+\displaystyle\frac{2\kappa\tau}{m}\,(2r-d)}
\label{eq:a3}
\end{equation}

\noindent and therefore we expect the pedestrian movement to become stationary 
after this time. It can be easily checked that $t_c$ drops to less than 0.1~s 
for compression distances as small as 1~mm. This means that the moving 
pedestrian's velocity will be close to the stationary velocity if the passing 
through process scales to $t\gg t_c$. \\

The stationary velocity $v_\infty$ can be obtained from Eq.~(\ref{eq:a1}) and 
the condition $\dot{v}=0$. Thus,

\begin{equation}
v_\infty=t_c\,\bigg[\displaystyle\frac{f_s}{m}+\displaystyle\frac{v_d}{\tau}
\bigg]\label{eq:a4}
\end{equation}

This is (approximately) the velocity that the moving pedestrian will 
hold most of the time while trying to get released from the other individuals. 
Thus, the time delay $t_d$ while passing across the still pedestrians 
will scale as $v_\infty^{-1}$. \\
 
Notice from Eqs.~(\ref{eq:a3}) and (\ref{eq:a4}) that $v_\infty$ decreases for 
increasing compression values. Also, an increase in the values of $f_s$ or 
$v_d$ will cause the corresponding increase in $v_\infty$. The resulting value 
for $v_\infty$ is a balance between the distance $2r-d$ and the forces $f_s$ or 
$v_d$. The distance $2r-d$, however, resembles the compression between members 
of the same blocking cluster, while the force $f_s$ corresponds to individuals 
out of the blocking cluster. \\

\subsection{The force balance} 

Fig.~\ref{fig:2} shows a schematic diagram for the forces applied to one of the 
still individuals. The force $f$ in Fig.~\ref{fig:2} represents the repulsive 
feeling actuating on the still individual due to the moving pedestrian. The 
force $\mathcal{F}$ is the required counter force necessary to keep the 
individual still. That is, $\mathcal{F}$ balances the repulsive feeling $f$ for 
a specific compression distance $2r-d$ (and fix values of $f_s$ and $v_d$).  
According to Section \ref{sfm}, the relationship between the compression 
distance and $\mathcal{F}$ (or $f$) is as follows

\begin{equation}
2r-d=B\,\ln(\mathcal{F}/A)\label{eq:a5}
\end{equation}

\noindent for the known values $A$ and $B$. \\

The relation (\ref{eq:a5}) can be applied to the expression (\ref{eq:a3}) for 
computing the characteristic time $t_c$. This means that $t_c$ may be 
controlled by $\mathcal{F}$, and consequently, it controls the stationary 
velocity $v_\infty$, according to (\ref{eq:a4}). Actually, the value of 
$v_\infty$ results from the balance between  $\mathcal{F}$ and $f_s$ (and 
$v_d$).   \\  

\subsection{\label{context}The crowd context} 

The above relations for a single moving pedestrian sliding between two still 
individuals should be put in the context of an evacuation process. These three 
pedestrian may belong to a ``blocking structure'', as defined in Section 
\ref{human}. The blocking structure may be surrounded by a large number of 
pedestrians that do not belong to this structure, but continuously pushes the 
structure towards the exit. Therefore, the forces $f_s$ and $\mathcal{F}$ are 
similar in nature and somehow represent the pressure actuating on the blocking 
structure from the surrounding crowd. \\

The pressure from the crowd depends on the anxiety level of the pedestrians. It 
has been shown that, at equilibrium, the crowd pressure grows linearly with the 
desired velocity $v_d$ and the number of individuals pushing from behind 
(see Ref.~\cite{Dorso5}). It seems reasonable, as a first approach, that $f_s$ 
and $\mathcal{F}$ varies as $\beta v_d$ for any fixed coefficient $\beta$.  \\

\begin{figure}
\includegraphics[width=\columnwidth]{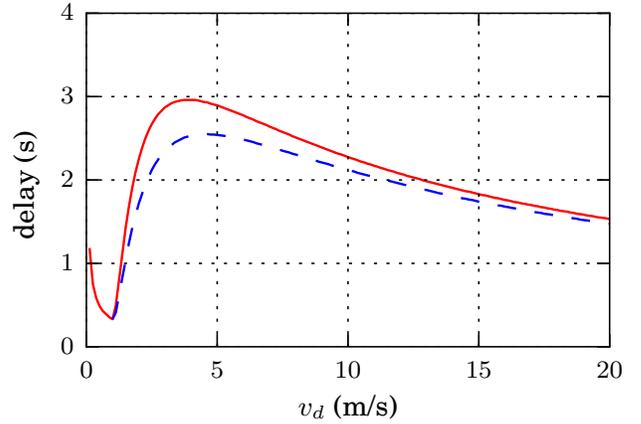}
\caption{\label{fig:3} (Color on-line only) Time delay ($v_\infty^{-1}$) for a 
moving individual passing between two still pedestrians, as shown in 
Fig.~\ref{fig:1and2}. The time interval was measured along 10~m across the still 
pedestrians. The initial velocity was $v_d$. The continuous line corresponds to 
the measured delay for $\beta v_d=2000\,v_d$ and $f=A\,\exp[(2r-d)/B]$ (see text 
for details). The dashed line corresponds to the measured delay for  
$\beta v_d=2000\,v_d$ and $f=A\,\exp[(2r-d)/B]+k\,(2r-d)$ (see 
text for details). The minimum time delay for both lines takes place at 
$v_d=1\,$m/s. The maximum time delay for the continuous line takes place at 
$v_d=3.7\,$m/s, while for the dashed line takes place at $v_d=4.2\,$m/s.  }
\end{figure}

The forces $f_s$ and $\mathcal{F}$ may be replaced by $\beta v_d$ in 
Eq.~(\ref{eq:a4}) for the evacuation process scenario, as explained in 
Section~\ref{stationary}. Thus, the stationary velocity $v_\infty$ only depends 
on the desired velocity of the pedestrians (and the total number of 
individuals). Fig.~\ref{fig:3} shows the behavior of the time delay 
($v_\infty^{-1}$) for a wide range of desired velocities $v_d$. \\ 

The continuous line in Fig.~\ref{fig:3} exhibits a local minimum and a maximum 
at $v_d=1\,$m/s and $v_d=3.7\,$m/s, respectively. The behavioral pattern for 
$v_d<1\,$m/s corresponds to non-contacting situations (that is, $2r-d<0$). 
The characteristic time for this regime is $t_c=\tau$, and thus, the time delay 
decreases for increasing values of $v_d$, according to Eq.~(\ref{eq:a4}). \\

The regime for $v_d>1\,$m/s corresponds to those situations where the moving 
pedestrian gets in contact with the two still individuals. Since the 
compression distance $2r-d$ becomes positive, there is a reduction in the  
characteristic time $t_c$, according to Eq.~(\ref{eq:a3}). This reduction 
actually changes the value of the stationary velocity $v_\infty$, as expressed 
in (\ref{eq:a4}). It is not immediate whether the $t_c$ reduction increases or 
decreases the velocity $v_\infty$. A closer inspection of the $v_\infty$ 
behavioral pattern is required.  \\

The computation of the slope for $v_\infty$ with respect to $v_d$ gives the 
following expression

\begin{equation}
\displaystyle\frac{dv_\infty}{dv_d}=\bigg[1-\displaystyle\frac{2\kappa 
B}{m}\,t_c\bigg]\,\displaystyle\frac {v_\infty}{v_d}\label{eq:a6}
\end{equation}

This expression shows a change of sign in the slope of $v_\infty$ for 
increasing values of $v_d$. It can be checked over that the expression enclosed 
in brackets is negative for small compressions, but as $t_c$ decreases 
(due to $v_d$ increments), it becomes positive. The vanishing condition for 
(\ref{eq:a6}) is  

\begin{equation}
B\ln\bigg(\displaystyle\frac{\beta 
v_d}{A}\bigg)=B-\displaystyle\frac{m}{2\kappa\tau 
}\label{eq:a7}
\end{equation}

The last term on the right becomes neglectable with respect to $B$ for the 
current literature values. Thus, the maximum time delay ($v_\infty^{-1}$) takes 
place close to $v_d=2.7\,A/\beta$. The corresponding compression distance for 
this desired velocity is $2r-d=B$. \\

\subsection{\label{appendix_remaks}Remarks}

The above computations show two relevant $v_d$ values: the one where a 
minimum time delay takes place and the one where the maximum time delay 
happens. The former corresponds to  $v_d=A/\beta$, or equivalently, $2r-d=0$. 
The latter corresponds to $v_d=2.7\,A/\beta$ or $2r-d=B$ (approximately).  \\  

The forces $f_s$ and $\mathcal{F}$ are similar in nature for the 
evacuation scenario. Therefore, $\mathcal{F}$ can be replaced by $f_s$ in the 
Eq.~(\ref{eq:a5}) for the stationary passing through process shown in 
Fig.~\ref{fig:1and2}. The stationary balance for Eq.~(\ref{eq:a1}) then reads

\begin{equation} 
A\,e^{(2r-d)/B}+\displaystyle\frac{mv_d}{\tau}=\bigg[2\,\kappa\,
(2r-d)+\displaystyle\frac{m}{\tau}\bigg]\,v_\infty\label{eq:a8}
\end{equation}

Accordingly, the time delay reads

\begin{equation}
 v_\infty^{-1}=\displaystyle\frac{1+\displaystyle\frac{2\kappa\tau}{m}\,
(2r-d)}{\displaystyle\frac{A\tau}{m}\,e^{(2r-d)/B}+v_d}\label{eq:a9}
\end{equation}

Notice from this expression that small increments of $2r-d$ produce 
increasing values of the time delay $v_\infty^{-1}$ if $2r-d<B$. But, further 
compression increments (that is, increments beyond $2r-d>B$) reduce the time 
delay, since the exponential function grows increasingly fast. \\

The above observations give a better understanding for the local maximum 
exhibited in Fig.~\ref{fig:3}. The positive slope range for 
$v_\infty^{-1}$ corresponds to small values of $f_s$ (that is, small values for 
the exponential function in (\ref{eq:a9})), while the negative slope range 
(beyond the local maximum) corresponds to high $f_s$ values.   \\

Although Fig.~\ref{fig:3} is in correspondence with Eq.~(\ref{eq:a6}), the 
local maximum does not actually take place at $v_d=2.7\,$m/s but at  
$v_d=3.7\,$m/s. This is right since Fig.~\ref{fig:3}  represents a complete 
simulation of the moving pedestrian instead of the stationary model for the 
pedestrian at the crossing point between the still individuals, as expressed in 
Eq.~(\ref{eq:a1}) and shown in Fig.~\ref{fig:1and2}.\\

Fig.~\ref{fig:3} also shows in dashed line the time delay for individuals with 
non-neglectable elastic compressions (see caption for details). The local 
maximum also appears but for lower time delay values. \\





\newpage 
\bibliography{paper}

\end{document}